\documentclass[11pt]{article}

\usepackage{amsmath}
\usepackage{amssymb}
\usepackage{epsfig}

\newcommand{\bigR}{{\mathbb{R}}}

\newcommand{\bigZ}{{\mathbb{Z}}}

\newcommand{\spacing}[1]{\renewcommand{\baselinestretch}{#1}\large\normalsize}

\newcommand{\vT}{\varepsilon T}

\newtheorem{dfn}{Definition}[section]
\newtheorem{rmk}[dfn]{Remark}
\newtheorem{thm}[dfn]{Theorem}

\newtheorem{lem}[dfn]{Lemma}

\begin{document}

\newpage

\title{Sufficient Conditions \\for \\ Fast Switching Synchronization \\ in \\ Time Varying Network Topologies
\author{Daniel J. Stilwell$^1$,  Erik M. Bollt$^2$, D. Gray Roberson$^1$\\
\vspace*{.1in}\\
\small $^1$Bradley Department of Electrical and Computer Engineering\\
\small Virginia Polytechnic Institute and State University\\
\small Blacksburg, VA \\
\vspace*{.1in} \\
\small $^2$Department of Mathematics and Computer Science \\
\small Clarkson Unviversity\\
\small Potsdam, NY}}
\maketitle

\begin{abstract}
In previous work \cite{skufca-bollt}, empirical evidence indicated that a time-varying network could propagate sufficient information to allow synchronization of the sometimes coupled oscillators, despite an instantaneously disconnected topology.
We prove here that if the network of oscillators synchronizes for the static time-average of the topology, then the network will synchronize with the time-varying topology if the time-average is achieved sufficiently fast.  Fast switching, fast on the time-scale of the coupled oscillators, overcomes the descychnronizing decoherence suggested by  disconnected instantaneous networks.  This result agrees in spirit with that of \cite{skufca-bollt} where empirical evidence suggested that a moving averaged graph Laplacian could  be used in the master-stability function analysis \cite{pecora.carroll.PRL.98}.   A new fast switching stability criterion here-in gives sufficiency of a fast-switching network leading to synchronization. Although this sufficient condition appears to be very conservative, it provides new insights about the requirements for synchronization when the network topology is time-varying.  In particular, it can be shown that networks of oscillators can synchronize even if at every point in time the frozen-time network topology is insufficiently connected to achieve synchronization.  
\end{abstract}

MCS: {37A25, 37D45, 37N25, 93B05, 94C15, 05C80, 92D30} 


\newpage
\spacing{1.25}
\section{Introduction}

Since Huygen's early observations of  weakly coupled clock pendula \cite{Huygens}, synchronization has been found  in a wide variety of phenomena, ranging from biological systems that include fire flies in the forest \cite{flies, strogatz1}, animal gates \cite{gates}, descriptions of the heart \cite{heart1, heart2, heart3}, and improved understanding of brain seizures \cite{schiff}, to chemistry \cite{chemical},  nonlinear optics \cite{Roy}, and  meteorology \cite{meteorology}.  See one of the many excellent reviews now available, including \cite{review1, review2, review3, strogatz2, Winfree, postnov}.  In particular, it has been known for more than 20 years that chaotic oscillators can synchronize under suitable coupling \cite{old1, old2, old3}.
Meanwhile, in recent years, the study of large scale, random networks has 
become an extremely active area with the advent of  advances in both theory and scientific application across numerous disciplines, as reviewed in \cite{reviewn1, reviewn2, reviewn3, reviewn4, reviewn5} for example. 
Recent investigations have sought to characterize how oscillator elements coupled according to a large-scale network architecture are impacted by the choice of architecture and corresponding spectral properties of the network, \cite{net1, net2, net3, net4, net5, net6}.   In particular, the master-stability function formalism \cite{pecora.carroll.PRL.98, pecora.etal.IJBC.2000} relates spectral properties of the graph Laplacian of the network to synchrony of supported oscillators, and this has been used in study of synchronization stability on arbitrary network architecture, \cite{net1}.  

Despite the very large literature to be found, the great majority of research activities have been focused on static networks  whose connectivity and coupling strengths are constant in time.  For example, static networks are assumed for the analysis of \cite{pecora.carroll.PRL.98, pecora.etal.IJBC.2000, net1}.  However, there are applications where the coupling strengths and even the network topology can evolve in time.  Recent work such as \cite{time1,time2,time3} are amongst the few to consider time-dependent couplings.  See also \cite{Kataoka} in which a so-called ``function dynamics" gives rise to networks that evolve according to a dynamical system, somewhat similarly to our networks.  In prior work \cite{skufca-bollt}, we were motivated by applications that include how a disease might occur in a network of agents in which the agents move, but the disease itself has its own time-scale.  We describe a competition of two time-scales.  Said plainly, the  disease has a natural typical incubation rate, and a natural infections rate (for example in a Susceptible$\rightarrow$Exposed$\rightarrow$Infected$\rightarrow$Recovered, SEIR, model), so if a susceptible agent does not come in contact  with an infected agent in the disease time scale, then there should be no new infection.
Mathematically, we constructed a ``moving neighborhood network" (MNN), a network of agents which move ergodically, and connect when in close proximity to each other.  Such a network was shown to lead to a Poisson distributed degree distribution, instantaneously, and hence the network was typically instantaneously disconnected.  It consists of typically many small subcomponents at each instant.  Such a description alone would suggest that there could be no global synchronization of the oscillators carried by each agent which are coupled according to the disconnnected network.  However, it was found that if the agents move quickly enough, then roughly described, in any recent time window, a given agent might be likely to have had some amount of coupling to most other agents.  It turns out that for fast enough moving agents,  these random time varying connections were enough to overcome even chaotic oscillators' sensitive dependence tendency to drift apart asynchronysly.  We formalized this idea by introducing a new description of the connectivity, a ``moving averaged" graph Laplacian.  We showed empirically that the spectrum of this construction works quite well together with the master stability formalism to accurately predict synchronization.

Besides our original motivation in mathematical epidemiology, it can be argued that this work has strong connections to ad hoc communication systems and control systems on time varying networks.  Fundamental connections between chaotic oscillations and proof of synchronization through symbolic dynamics \cite{syncsymb, ned1} and control  \cite{ned2, me1, Hayes} and even definition of chaos through symbolic dynamics suggest this work is rooted in a description of information flow in the network.

Coordinated control for platoons of autonomous vehicles can also be addressed using network concepts \cite{desai.etal.TRA.2001,roberson.stilwell.ACC.05,fax.murry.TAC.2004}.  Each vehicle is represented by a node, and communication or mutual sensing is represented by connections between nodes.  In \cite{fax.murry.TAC.2004} the average position of a platoon of vehicles is regulated, and the graph Laplacian is used to describe communication between vehicles.  It is shown that the spectrum of the graph Laplacian can be used to indicate stability of the controlled system.  As pointed out in \cite{roberson.stilwell.ACC.05}, the use of a graph Laplacian is not entirely common since it appears naturally for only a limited class of control objectives.

These considerations have led us in this work to consider a simplified version of the moving agents of our MNN model.  Considering certain time varying coupled network architectures, we can now make rigorous but sufficient statements concerning fast switching, and we use mathematical machinery not so far typically used in the synchronization community.
The main result of this work comes from the fields of switched systems, and specifically builds on the concept of \emph{fast switching}.  Switched systems are a class of systems whose coefficients undergo abrupt change, for example, consider the linear state equation
\begin{equation} \label{eq.A.example}
    \dot{x}(t) = A_{\rho(t)} x(t)
\end{equation}
where $\rho(t): \bigR \mapsto \bigZ_+$ is a switching sequence that selects elements from a family of matrix-valued coefficients $\Theta = \{ A_1, A_2, \ldots \}$.  When each element of $\Theta$ is Hurwitz, stability of \eqref{eq.A.example} is guaranteed if $\rho(t)$ switches sufficiently slowly.  Further restrictions on elements of $\Theta$, such as existence of a common Lyapunov function, can guarantee stability for arbitrary switching functions, including those that are not slow.  An excellent overview of the field of switched systems and control is presented in \cite{liberzon.morse.csm1999} and in the book \cite{liberzon.book.2003}.

Even when the elements of $\Theta$ are not all Hurwitz, stability of \eqref{eq.A.example} is still possible, although the class of switching functions is further restricted.  For example, in \cite{wicks.etal.EJC.1998} a stabilizing switching sequence is determined by selecting elements of $\Theta$ based on the location of the state $x(t)$ in the state space.  This is essentially a form of state feedback.

When no elements of $\Theta$ are Hurwitz, which is the case that is considered herein, stability of \eqref{eq.A.example} can sometimes be guaranteed if the switching sequence is sufficiently fast.  Loosely speaking, it can be shown that
\begin{equation} \label{eq.A.eps}
    \dot{x}(t) = A_{\rho(t/\varepsilon)}x(t)
\end{equation}
is asymptotically stable if the time-average 
\[
    \frac{1}{T} \int_t^{t+T} A_{\rho(\sigma)} d\sigma
\]
is Hurwitz for all $t$, and if $\varepsilon$ is sufficiently small.  This fact has been established in \cite{bentsman.etal.TAC.1987,bellman.etal.TAC.1985,tokarzewsi.IJSC.1987} for several classes of linear systems related to \eqref{eq.A.eps}.  Similar results have been presented in \cite{aeyels.peuteman.automatica.1999,aeyels.peuteman.TAC.1998} for classes of nonautonomous nonlinear systems where time is parameterized by $t/\varepsilon$ as in \eqref{eq.A.eps}.  In this case, stability of a specific average system implies stability of the original system if $\varepsilon$ is sufficiently small.  In addition, this work requires the existence of a Lyapunov function that is related to a certain average of the system but which is not a function of time.  This requirement is too restrictive for the class of linear time-varying systems considered herein.   A new fast switching stability condition, presented in Section \ref{sec.main.result}, is derived in order to assess local stability of networked oscillators about the synchronization manifold.

\section{Preliminaries} \label{sec.preliminaries}
We consider a network of coupled oscillators consisting of $r$ identical oscillators,
\begin{equation} \label{eq.single}
    \dot{x}_i(t) = f(x_i(t)) + \sigma B \sum_{j=1}^{r} l_{ij}(t) x_j(t), \quad i=1, \ldots, r 
\end{equation}
where $x_i(t)\in\bigR^n$ is the state of oscillator $i$, $B\in\bigR^{n\times n}$, and the scalar $\sigma$ is a control variable that sets the coupling strength between oscillators.  The scalars $l_{ij}(t)$ are elements of the graph Laplacian of the network graph and describe the interconnections between individual oscillators.  Let $G(t)$ be the time-varying graph consisting of $r$ vertices $v_i$ together with a set of ordered pairs of vertices $\{v_i, v_j\}$ that define the edges of the graph.   In this work, we assume that $\{v_i, v_i\}\in G(t)$ for $i=1, \ldots, r$.  Let $\tilde{G}(t)$ be the $r \times r$ adjacency matrix corresponding to $G(t)$, then $\tilde{G}_{i,j}(t) = 1$ if $\{v_i, v_j\}$ is an edge of the graph at time $t$ and $\tilde{G}_{i,j}(t) = 0$ otherwise.  The graph Laplacian is defined as 
\begin{equation} \label{eq.define.L}
    L(t) = \operatorname{diag}( d(t)) - \tilde{G}(t) 
\end{equation}
where the $i^{\text{th}}$ element of $d(t)\in\bigR^{r}$ is the number of vertices that vertex $i$ is connected to, including itself. 

Synchronization can be assessed by examining local asymptotic stability of the oscillators along the synchronization manifold.  Linearizing each oscillator \eqref{eq.single} about the trajectory $x^o(t)$, which is assumed to be on the synchronization manifold, yields
\begin{equation} \label{eq.linear.single}
    \dot{z}_i(t) = F(t)z_i(t) + \sigma B \sum_{j=1}^{r} l_{ij}(t)z_j(t)
\end{equation}
where $z_i(t) = x_i(t) - x^o(t)$, and $F(t)= Df$ evaluated at $x^o(t)$. Let $L(t)$ be the $r\times r$ matrix composed of entries $l_{ij}(t)$, then the system of linearized coupled oscillators is written
\begin{equation} \label{eq.lin1}
\begin{split}
    \dot{z}(t) &=\left(I_r \otimes F(t) + \sigma (I_n \otimes B) (L \otimes I_r)\right) z(t)\\
        &  =\left(I_r \otimes F(t) + \sigma L\otimes  B\right) z(t)\\
\end{split}
\end{equation}
where `$\otimes$' is the Kronecker product and $z(t) = [ z_1^T(t), \ldots, z_r^T(t)]^T$.  Standard properties of the Kronecker product are utilized here and in the sequel, including: for conformable matrices $A$, $B$, $C$, and $D$, $(A\otimes B)(C\otimes D) = AC\otimes BD$.  Notation throughout is standard, and we assume that $\|\cdot \|$ refers to an induced norm.

It has been shown in \cite{pecora.carroll.PRL.98,pecora.etal.IJBC.2000} that the linearized set of oscillators \eqref{eq.lin1} can be decomposed into two components: one that evolves along the synchronization manifold and another that evolves transverse to the synchronization manifold.  If the latter component is asymptotically stable, then the set of oscillators will synchronize.

The claimed decomposition is achieved using a Schur transformation.  We briefly describe the decomposition since it plays a central role in our assessment of synchronization under time-varying network connections.  Let $P\in\bigR^{n\times n}$ be a unitary matrix such that $U = P^{-1}LP$ where $U$ is upper traigular.  The eigenvalues $\lambda_1, \ldots, \lambda_r$ of $L$ appear on the main diagonal of $U$. Then a change of variables $\xi(t) = \left(P \otimes I_n\right)^{-1}z(t)$ yields

\begin{equation} \label{eq.linT1}
\begin{split}
    \dot{\xi}(t) & = \left(P \otimes I_n\right)^{-1}\left(I_r \otimes F(t) + \sigma L\otimes B\right) \left(P \otimes I_n\right)\xi(t) \\
     & = \left(I_r \otimes F(t) + \sigma P^{-1}LP\otimes B\right)\xi(t) \\
     & = \left(I_r \otimes F(t) + \sigma U\otimes B\right)\xi(t) \\
\end{split}
\end{equation}
Due to the block-diagonal structure of $I_r \otimes F(t)$ and the upper triangular structure of $U$, stability of \eqref{eq.linT1} is equivalent to stability of the subsystems,
\begin{equation} \label{eq.lam1}
   \dot{\xi}_i(t) = (F(t) + \sigma \lambda_iB) \xi_i(t), \qquad i=1, \ldots, r
\end{equation}
where $\lambda_1, \ldots, \lambda_r$ are the eigenvalues of $L$.  Note that since the row sums of $L$ are zero, the spectrum of $L$ contains at least one zero eigenvalue, which we assign $\lambda_1=0$. Thus
\[
    \dot{\xi}_1(t) = F(t) \xi_1(t)
\]
evolves on the along the synchronization manifold, while \eqref{eq.lam1} with $i=2, \ldots, r$ evolves transverse to the synchronization manifold \cite{pecora.carroll.PRL.98}.  Since the oscillators are assumed identical, the (identity) synchronization manifold  is invariant for all couplings, the question being its stability.  The set of coupled oscillators will synchronize if the synchronization manifold is stable, if \eqref{eq.lam1} with $i=2, \ldots, r$ is asymptotically stable.

\section{Main Result} \label{sec.main.result}
For a given static network, the master stability function characterizes values of $\sigma$ for which a set of coupled oscillators \eqref{eq.lin1} synchronizes \cite{pecora.carroll.PRL.98,Barahona,Fink}. The graph Laplacian matrix $L$ has $n$ eigenvalues, which we label,
\begin{equation}
0=\lambda_0 \leq \ldots \leq \lambda_{n-1}=\lambda_{max}.
\end{equation}  
The stability question reduces by linear perturbation analysis to a constraint upon the eigenvalues of the Laplacian:  
\begin{equation}
\sigma \lambda_i \in (\alpha_1,\alpha_2) \quad \forall i=1, \ldots , n-1,
\label{estaba}
\end{equation}
where $\alpha_1,\alpha_2$ are given by the {\it master stability function} (MSF), a property of the oscillator equations.  For $\sigma$ small, synchronization is unstable if $\sigma \lambda_1 < \alpha_1;$ as $\sigma$ is increased, instability arises when,
\begin{equation}
\sigma \lambda_{max} > \alpha_2.
\end{equation}
   By algebraic manipulation of (\ref{estaba}),  if,
\begin{equation}
\frac{\lambda_{max}}{\lambda_1} < \frac{\alpha_2}{\alpha_1}=:\beta,
\end{equation}
 then there is a coupling parameter, $\sigma_s,$ that will stabilize the synchronized state.  For some networks, no value of $\sigma$ satisfies (\ref{estaba}).  In particular, since the multiplicity of the zero eigenvalue defines the number of completely reducible subcomponents, if $\lambda_1=0,$ the network is not connected, and synchronization is not stable.  However, even when $\lambda_1>0,$ if the spread of eigenvalues is too great, then synchronization may still not be achievable.

For the case of a time-varying network topology, represented by $L(t)$, our principal contribution is to show that the network can synchronize even if the static network $\bar{L} = L(t)$ for any frozen value of $t$ is insufficient to support synchronization.  Specifically, we show that the time-average of $L(t)$, not the frozen values of $L(t)$, is an indicator of synchronization. If the time-average of $L(t)$ is sufficient to support synchronization, then the time-varying network will synchronize if the time-average is achieved sufficiently fast.

\begin{thm} \label{thm.thm1}
Suppose a set of coupled oscillators with linearized dynamics 
\begin{equation} \label{eq.thm.Lbar}
    \dot{z}_s(t) = \left(I_r \otimes F(t) + \sigma  \bar{L}\otimes B\right) z_s(t)
\end{equation}
achieves synchronization.  Then there exists a positive scalar $\varepsilon^*$ such that the set of oscillators with linearized dynamics
\begin{equation} \label{eq.thm.Lt}
    \dot{z}_a(t) = \left(I_r \otimes F(t) + \sigma  L(t/\varepsilon)\otimes B\right) z_a(t)
\end{equation}
and time-varying network connections $L(t)$ achieves synchronization for all fixed $0 < \varepsilon < \varepsilon^*$ if $L(t)$ satisfies
\begin{equation} \label{eq.thm.Laverage}
   \frac{1}{T} \int_t^{t+T} L(\sigma) d\sigma = \bar{L}
\end{equation}    
and the column sums of $L(t)$ are all zero for all $t$.
\end{thm}

\begin{rmk}
Since $L(t)$ represents a time-varying network, we may assume that for each value of $t$, $L(t)$ is a graph Laplacian as defined in \eqref{eq.define.L}.  Thus the time-average $\bar{L}$ in \eqref{eq.thm.Laverage} is not a graph Laplacian. In other words, $\bar{L}$ does not necessarily correspond to a particular network topology and arises only as the time-average of $L(t)$.  However, $\bar{L}$ does inherit the zero row and column sum property of $L(t)$.
\end{rmk}

\noindent A preliminary lemma is required to prove Theorem \ref{thm.thm1},  the proof of which appears in the Appendix.

\begin{lem} \label{lem.prelim}
Suppose the matrix-valued function $E(t)$ is such that 
\begin{equation} \label{eq.E.average}
   \frac{1}{T} \int_t^{t+T} E(\sigma) d\sigma = \bar{E}
\end{equation}
for all $t$ and
\begin{equation} \label{eq.linE.bar}
    \dot{x}(t) = (A(t) + \bar{E})x(t), \qquad x(t_o) = x_o, \qquad t\geq t_o
\end{equation}
is uniformly exponentially stable.  Then there exists $\varepsilon^*>0$ such that for all fixed $\varepsilon \in (0, \varepsilon^*)$, 
\begin{equation} \label{eq.linE.t}
    \dot{z}(t) = (A(t)+E(t/\varepsilon))z(t), \qquad z(t_o) = z_o, \qquad t\geq t_o
\end{equation}
is uniformly exponentially stable.
\end{lem}

\noindent \emph{Proof of Theorem \ref{thm.thm1}}:\\
First we show that the Schur transformation that decomposes the set of oscillators \eqref{eq.thm.Lbar} with static $\bar{L}$ also induces a similar decomposition for \eqref{eq.thm.Lt} with time-varying $L(t)$.  Then we apply Lemma \ref{lem.prelim} to show that the modes of the system that evolve transverse to the synchronization manifold are stable if $\varepsilon$ is sufficiently small.

Let $P\in\bigR^{r\times r}$ be a unitary matrix such that $\bar{U}=P^{-1}\bar{L}P$ where 
\[
    \bar{U} = \begin{bmatrix} 
    0 & \bar{U}_{1} \\
    0_{(r-1)\times 1} & \bar{U}_{2}
    \end{bmatrix}
\]
is the Schur transformation of $\bar{L}$, and $\bar{U}_{2} \in\bigR^{(r-1)\times (r-1)}$ is upper triangular.  Without loss of generality, we have assumed that the left-most column of $P$ is the unity norm eigenvector $ \left[\sqrt{1/r}, \ldots, \sqrt{1/r}\right]^T$ corresponding to a zero eigenvalue.  The change of variables $\xi_s(t) = (P\otimes I)^{-1}z_s(t)$ yields the decomposition $\xi_s = [\xi_{s1}, \xi_{s2}]^T$ where $\xi_{s1}\in\bigR^n$, $\xi_{s2} \in \bigR^{n(r-1)}$, and $\xi_{s2}$ satisfies
\begin{equation}  \label{eq.s2}
    \dot{\xi}_{s2}(t) = \left(I_{r-1} \otimes F(t) + \sigma \bar{U}_{2}\otimes B \right) \xi_{s2}(t) 
\end{equation}
As discussed in Section \ref{sec.preliminaries}, \eqref{eq.s2} is asymptotically stable by hypothesis.

We now consider the same change of variables applied to \eqref{eq.thm.Lt}.  First, note that 
\[
    U(t) = P^{-1}L(t)P = 
    \begin{bmatrix} 
    0 & U_1(t) \\
    0_{(r-1)\times 1} & U_{2}(t)
    \end{bmatrix}
\]
since the column sums for $L(t)$ are zero for all $t$.  The change of variables $\xi_a(t) = (P\otimes I)^{-1}z_a(t)$ yields the decomposition $\xi_a = [\xi_{a1}, \xi_{a2}]^T$ where $\xi_{a1}\in\bigR^n$ evolves along the synchronization manifold and $\xi_{a2}\in \bigR^{n(r-1)}$ evolves transverse to the synchronization manifold.  To verify that the oscillators synchronize, it is sufficient to show that 
\begin{equation} \label{eq.a2}
    \dot{\xi_{a2}}(t) = \left(I_{r-1} \otimes F(t)+ \sigma  U_{2}(t/\varepsilon)\otimes B\right) \xi_{a2}(t)
\end{equation}
is asymptotically stable when $\varepsilon$ is sufficiently small.  Since 
\begin{align*}
      \bar{U} & = P^{-1}\bar{L}P \\
       &= \frac{1}{T} \int_t^{t+T} P^{-1}L(\sigma)P d\sigma  \\
       & = \frac{1}{T} \int_t^{t+T} U(\sigma) d\sigma  \\
 \end{align*}
 we conclude that 
 \begin{equation} \label{eq.Ubar}
      \bar{U}_2 = \frac{1}{T} \int_t^{t+T} U_2(\sigma) d\sigma  
 \end{equation}
 Thus the desired result is obtained by direct application of Lemma \ref{lem.prelim} along with \eqref{eq.s2}, \eqref{eq.a2}, and \eqref{eq.Ubar}.
\hfill $\blacksquare$

\section{Illustration}
To illustrate fast switching concepts applied to synchronization of a set of oscillators, we consider a set of $r$  Rossler attractors 
\begin{equation} \label{eq.rossler}
\begin{split}
    \dot{x}_i(t) &= -y_i(t) - z_i(t) - \sigma \sum _{j=1}^{r}l_{ij}(t/\varepsilon) x_j(t) \\
    \dot{y}_i(t) &= x_i(t) + a y_i(t) \\
    \dot{z}_i(t) &= b + z_i(t)(x_i(t)-c)
\end{split}
\end{equation}
where $i=1, \ldots , r$, $a= 0.165$, $b = 0.2$, $c = 10$, and $\sigma = 0.3$.  Oscillators are coupled through the $x_i$ variables via $l_{ij}(t)$.  Coupling between subsystems (nodes) is defined by a time-varying graph $G(t)$, with corresponding adjacency matrix $\tilde{G}(t)$.  The graph Laplacian $L(t)$, with entries $l_{ij}(t)$ is defined as in \eqref{eq.define.L}.

\begin{figure}[h!]
\begin{tabular}{c c c}
    \epsfig{file = 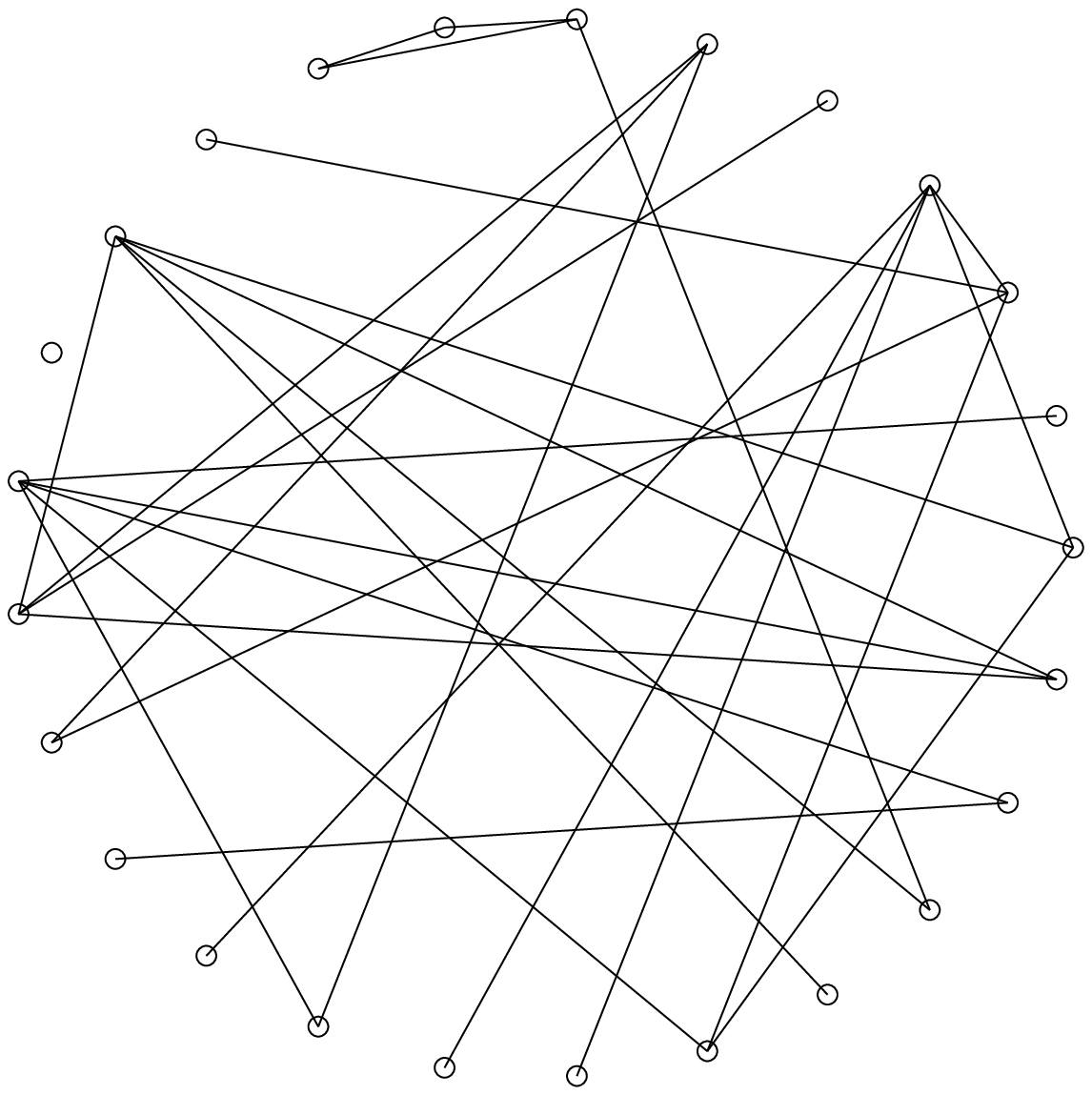, width = 2.2in} &
    \epsfig{file = 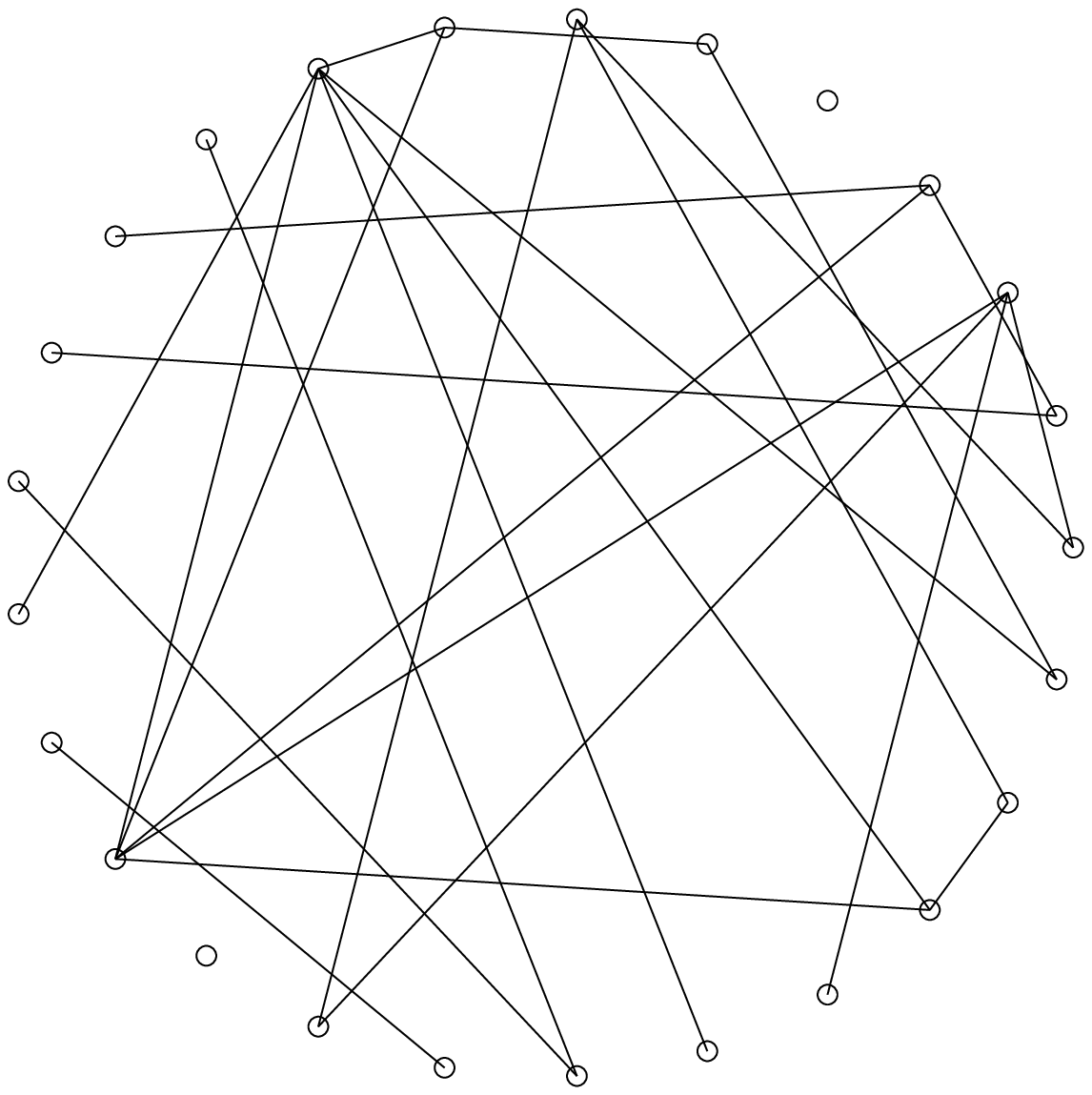, width = 2.2in} &
    \epsfig{file = 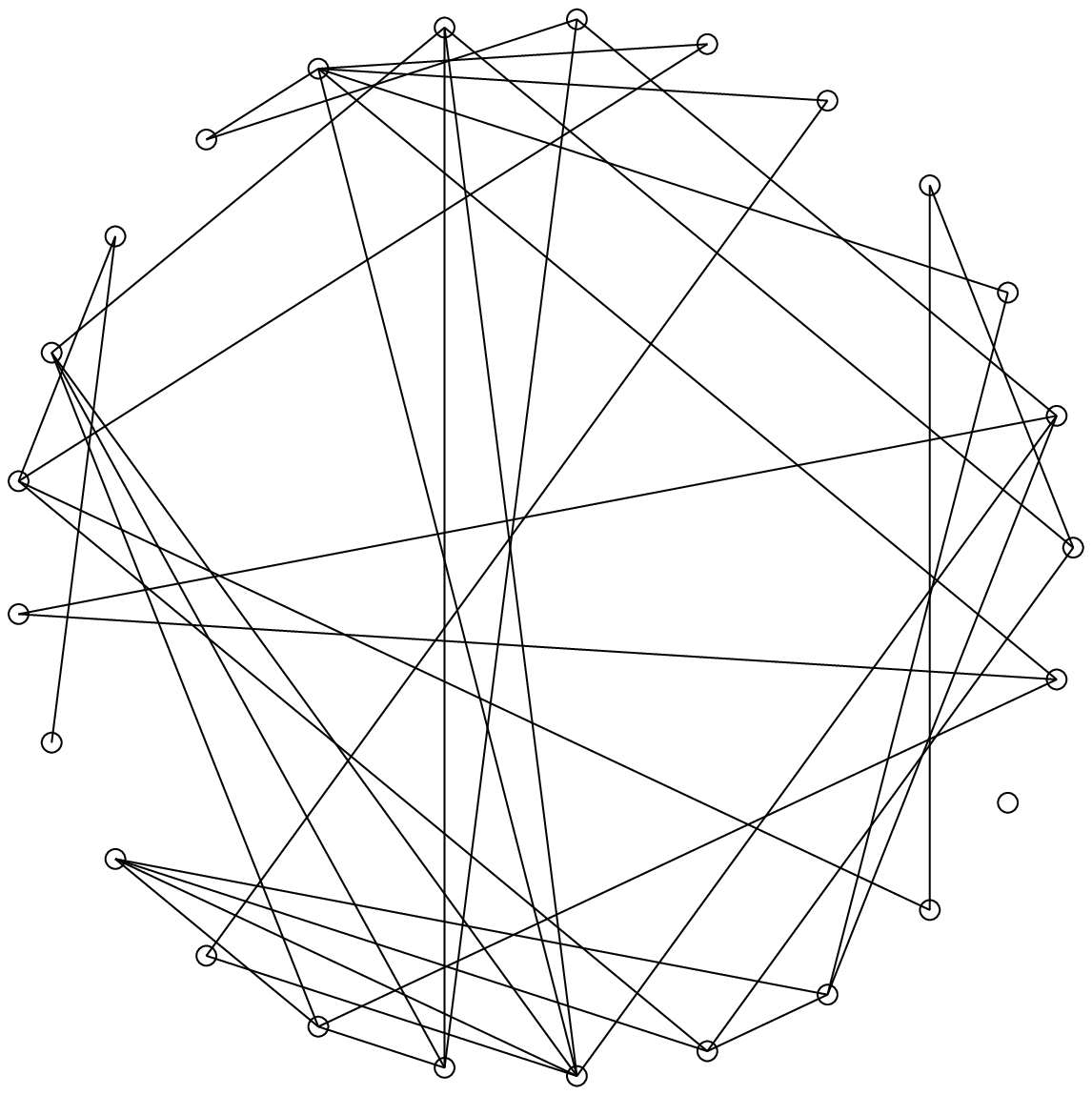, width = 2.2in} \\
    (a) & (b) & (c) \\
    \epsfig{file = 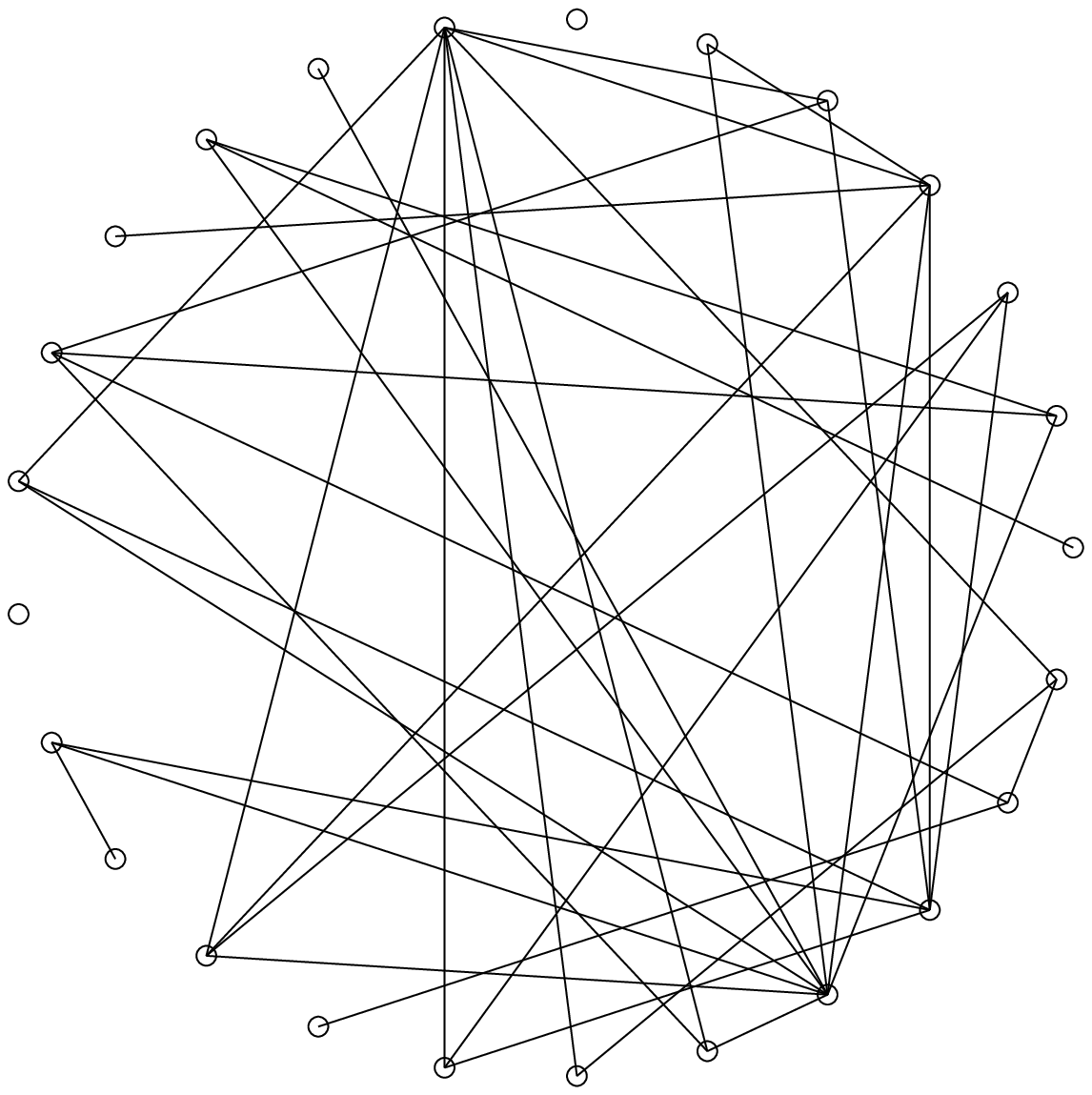, width = 2.2in} &
    \epsfig{file = 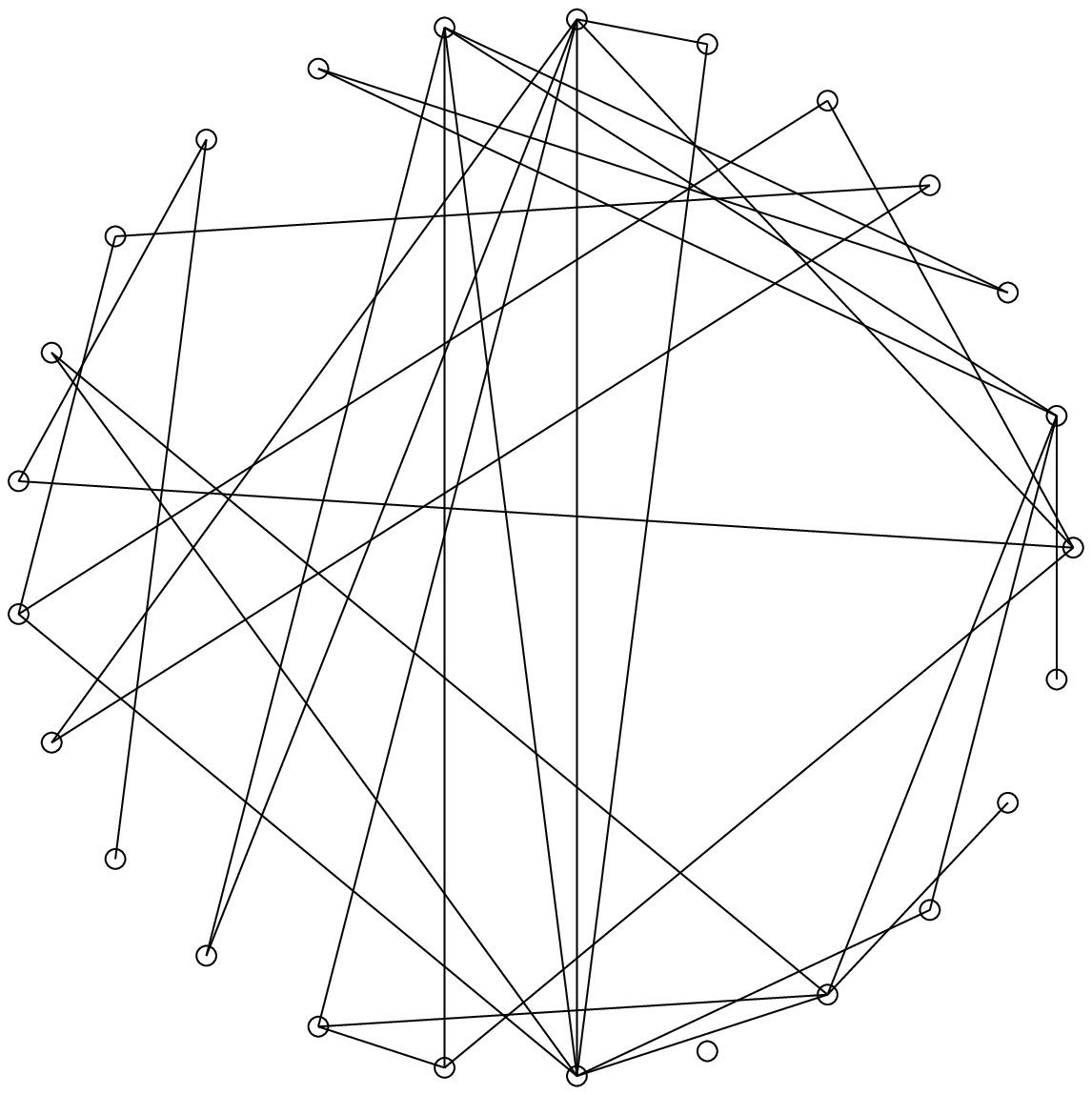, width = 2.2in} &
    \epsfig{file = 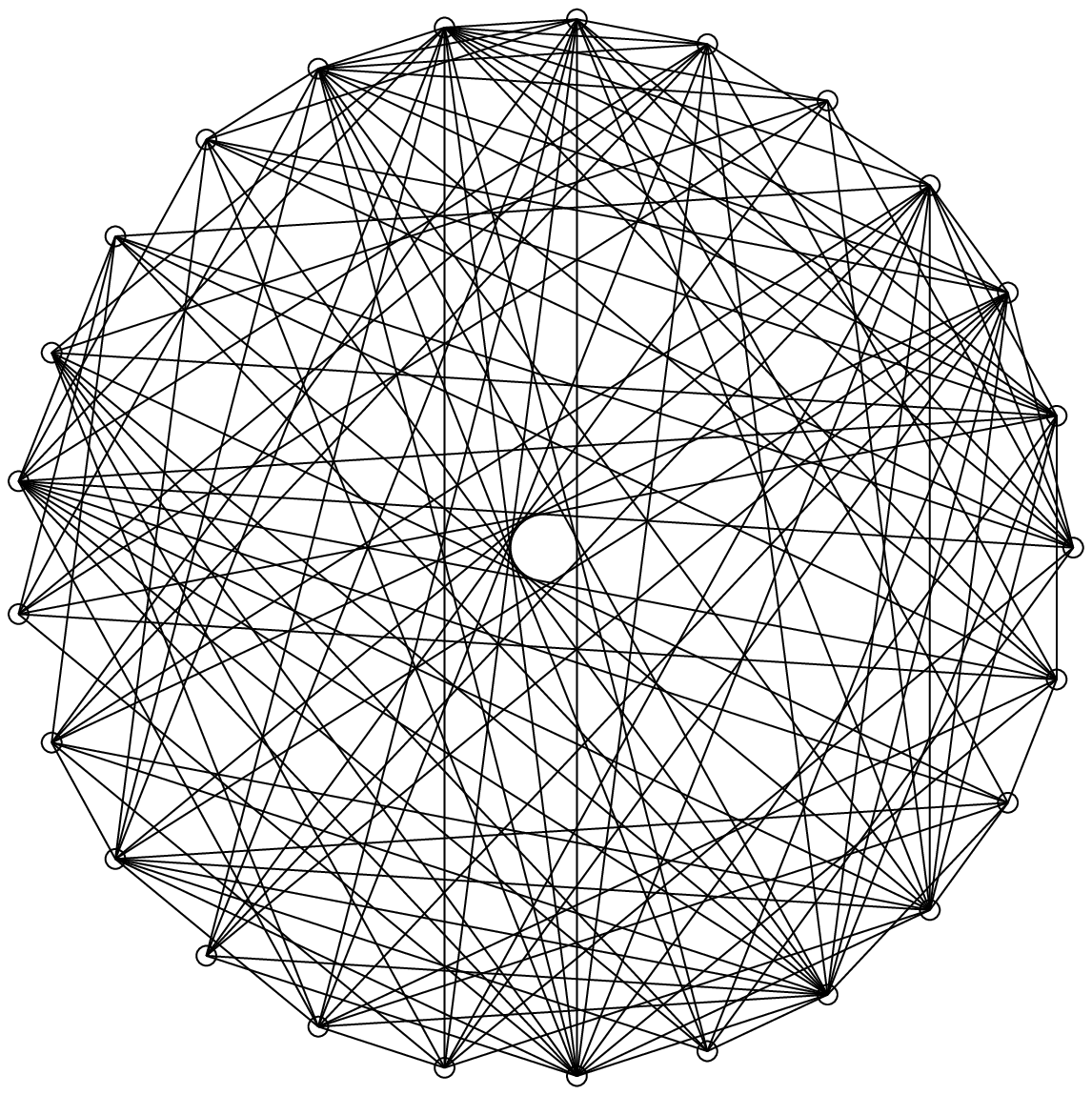, width = 2.2in} \\
        (d) & (e) & (f) \\
\end{tabular}
\caption{(a)-(e) are graphs $G_1$ through $G_5$, respectively, while (f) is the union of graphs. } \label{fig.all.graphs}
\end{figure}

For the purposes of illustration, we choose a set of five graphs and corresponding adjacency matrices $\tilde{G}_1, \ldots, \tilde{G}_5$ with the property that none of them are fully connected.  That is, each graph contains pairs of nodes that do not have a path between them.  However, the union of vertices over all five graphs yields a fully connected graph with the longest path  between nodes containing no more that two other nodes.  All five graphs and the union of graph vertices are shown in Figure \ref{fig.all.graphs}.  

A simple strategy is chosen for switching among graph Laplacians associated with the set of graphs.  We choose the $T$-periodic $L(t)$ defined over one period by
\[
    L(t) = \sum_{i=1}^5 L_i \chi_{[(i-1)T/5,\: iT/5)}(t)
\]
where $\chi_{[t_1, \: t_2)}(t)$ is the indicator function with support $[t_1, \: t_2)$.  The time-average of $L(t)$ is
\begin{equation} \label{eq.example.L.average}
\begin{split}
    \bar{L} &= \frac{1}{\varepsilon T} \int_0^{\varepsilon T }L(t/\varepsilon) dt \\
    & = \frac{1}{5}\sum_{i=1}^5 L_i
\end{split}
\end{equation}

\begin{figure}[h!]
    \centerline{\epsfig{file = 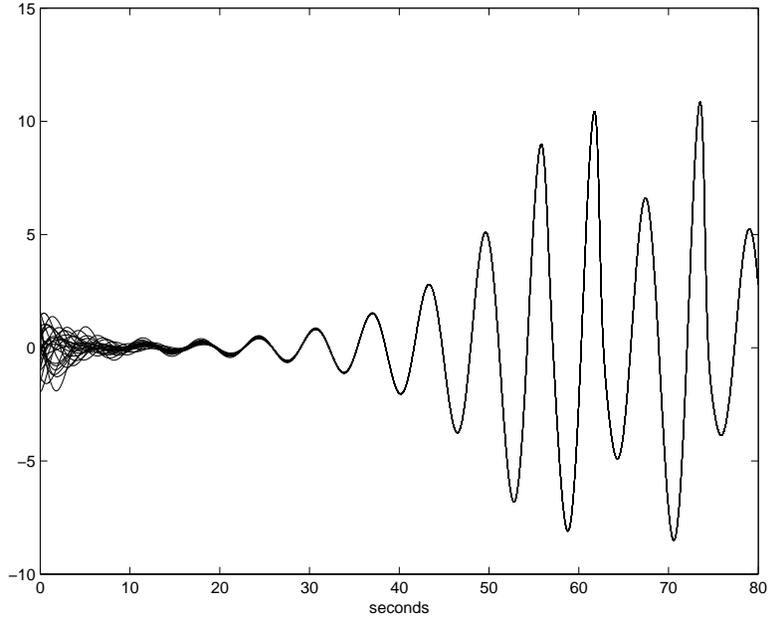, width = 4 in}}
    \caption{The $x$-coordinate for the set of coupled Rossler attractors using the average graph Laplacian.} \label{fig.average.oscillator}
\end{figure}

\begin{figure}[h!]
    \centerline{\epsfig{file = 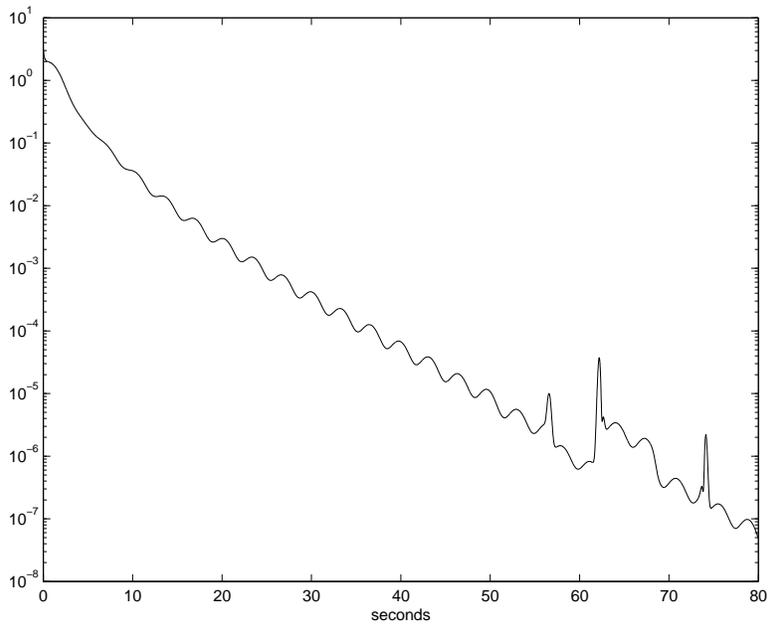, width = 4 in}}
    \caption{Root-mean-squared error in \eqref{eq.RMS} for the set of coupled Rossler attractors using the average network $\bar{L}$.} \label{fig.RMS.average}
\end{figure}

Toward computing the upper bound for $\varepsilon$ given by \eqref{eq.v2}, the set of coupled oscillators \eqref{eq.rossler}  with coupling defined by \eqref{eq.example.L.average} are integrated.  The $x$-coordinate for each oscillator is shown in Figure \ref{fig.average.oscillator}.  The oscillators clearly synchronize.  Asymptotic stability of the oscillators with respect to the synchronization manifold is suggested by plotting the sum-square deviation of the states
\begin{equation} \label{eq.RMS}
    \sum_{i=1}^{r} (x_i(t) - \mu_x(t))^2 +  (y_i(t) - \mu_y(t))^2 +  (z_i(t) - \mu_z(t))^2 
\end{equation}
about the averages
\[
    \mu_x(t) = \frac{1}{r} \sum_{i=1}^{r} x_i(t)
\]
where $\mu_y(t)$ and $\mu_z(t)$ are defined similarly.  Approximately exponential decay of \eqref{eq.RMS} is evident in Figure \ref{fig.RMS.average}.

The linear time-varying system \eqref{eq.lin1} corresponding to the set of coupled Rossler attractors is computed from the Jacobian of the right-hand side of \eqref{eq.rossler} evaluated at the solutions shown in Figure \ref{fig.average.oscillator}.

\begin{figure}[h!]
    \centerline{\epsfig{file = 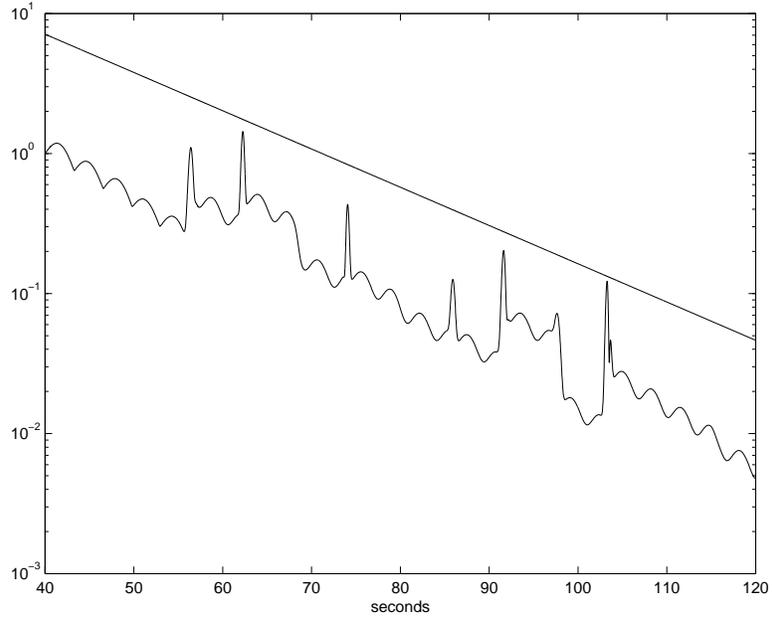, width = 4 in}}
    \caption{Norm of the transition matrix $\Phi(t,\tau)$ along with an exponentially decaying upper bound.} \label{fig.phi.bound}
\end{figure}

\begin{figure}[h!]
    \centerline{\epsfig{file = 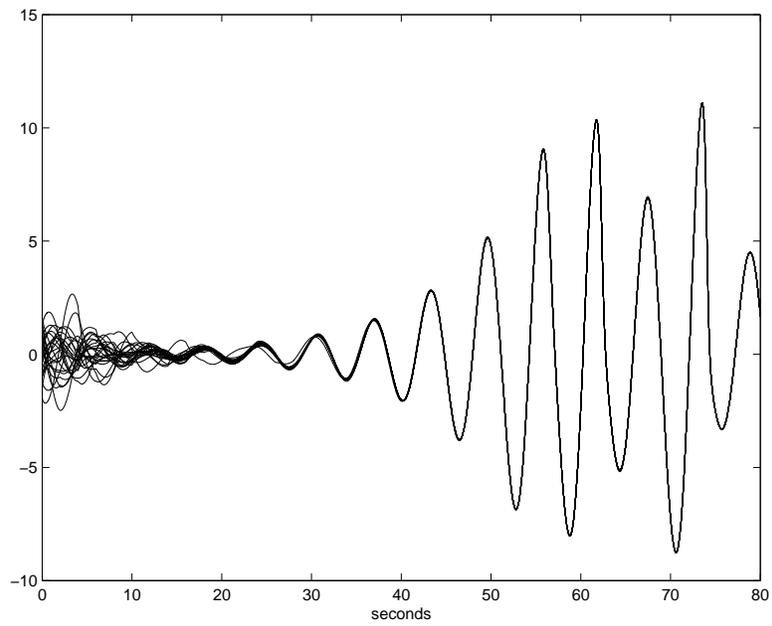, width = 4 in}}
    \caption{The $x$-coordinate for the set of coupled Rossler attractors using the switched network where $\varepsilon=1$.} \label{fig.oscillator.switched}
\end{figure}

As described in the proof of Lemma \ref{lem.prelim}, a Schur transformation $U$ that diagonalizes $\bar{L}$ is computed and used as a state transformation to decompose the linear time-varying system \eqref{eq.lin1} into a component that evolves along the synchronization manifold and another component that evolves transverse to the synchronization manifold. The upper bound for $\varepsilon$ given in Theorem \ref{thm.thm1} is computed from the component of the linear system that evolves transverse to the synchronization manifold,
\[
    \dot{\xi}_{a2}(t) = (I_{r-1}\otimes F(t) + \sigma U_2\otimes B) \xi_{a2}(t)
\]
We now estimate the constants $\alpha$, $\rho$, $\eta$, and $\mu$ needed to compute the right-hand side of \eqref{eq.v2} in the proof of Lemma \ref{lem.prelim} (see Appendix). This is used to compute an maximum value of $\varepsilon$.  The constant $\alpha$ is computed from \eqref{eq.define.alpha}, while the transition matrix is computed from 
\[
    \dot{\Phi}(t,\tau) = (I_{r-1}\otimes F(t) + \sigma U_2\otimes B)\Phi(t,\tau), \qquad \Phi(\tau, \tau) = I
\]
The norm of the transition matrix $\|\Phi(t,\tau)\|$ is shown in Figure \ref{fig.phi.bound}.  The initial time $\tau$ is chosen to be 40 seconds to ensure that the states of \eqref{eq.rossler} are reasonably close the the synchronization manifold.  An upper bound that satisfies $\|\Phi(t,\tau)\| \leq \gamma e^{-\lambda(t-\tau)}$ is also shown in Figure \ref{fig.phi.bound}.  The coefficients $\rho$, $\mu$ and $\eta$ in \eqref{eq.bound.phi} are computed from  $\gamma$ and $\lambda$ when evaluating the right-hand side of \eqref{eq.v2}.  Choosing $T=1$, the right-hand side of \eqref{eq.v2} is evaluated for this example, and we determine that the set of couple oscillators will synchronize if $\varepsilon < 3.3\times 10^{-7}$.  This shows that our bound is exceedingly conservative.  For example, empirically the oscillators will synchronize with $\varepsilon=1$, as shown in Figure \ref{fig.oscillator.switched}.


\section{Acknowledgments}

EMB has been supported by the National Science Foundation under  
DMS0404778.  DJS is support by the National Science Foundation under IIS0238092 and OCE0354810 and the Office of Naval Research under N00014-03-1-0444.


\clearpage
\newpage


\section{Appendix}
Proof of Lemma \ref{lem.prelim}

Since \eqref{eq.linE.bar} is uniformly exponentially stable, there exists a symmetric matrix function $Q(t)$ and positive scalars $\eta$, $\rho$, and $\mu$ such that the Lyapunov function 
\[
    v(x(t),t) = x^T(t)Q(t)x(t)
\]
satisfies
\begin{equation} \label{eq.v}
    \eta \|x(t)\|^2 \leq v(x(t),t) \leq \rho \|x(t)\|^2
\end{equation}
\begin{equation} \label{eq.vdot}
    \frac{d}{dt} v(x(t), t) \leq -\mu \|x(t)\|^2
\end{equation}
for all $t$.  To establish uniform exponential stability of \eqref{eq.linE.t}, we will show that  $v(z(t), t)$ is also a Lyapunov function for $\eqref{eq.linE.t}$ if $\varepsilon$ is sufficiently small.  This claim is achieved by showing that for sufficiently small values of $\varepsilon$, 
\begin{equation} \label{eq.deltav1}
    \Delta v(z, t+\vT, t) \equiv v(z(t+\vT), t+\vT) - v(z(t), t)
\end{equation}
is negative definite for all $t$.  Expanding \eqref{eq.deltav1} yields
\begin{align*}
    \Delta v(z, t+\vT, t) & = z^T(t+\vT)Q(t+\vT)z(t+\vT) - z^T(t)Q(t)z(t) \\
             & = z^T(t)\Big( \Phi^T_{E}(t + \varepsilon T , t) Q(\varepsilon T+t)\Phi_{E}(t + \varepsilon T, t) - Q(t) \Big) z(t) \\
\end{align*}
where $\Phi_{E}(t,\tau)$ is the transition matrix corresponding to $A(t) + E(t/\varepsilon)$.  Similarly denoting  the transition matrix for $A(t) + \bar{E}$ as $\Phi_{\bar{E}}(t,\tau)$, we define
\begin{equation*} 
\begin{split}
    H(t+ \vT,t) & = \Phi_{E}(t + \vT, t) - \Phi_{\bar{E}}(t + \vT, t) \\
            & = I + \int_t^{t + \vT} A(\sigma_1) + E(\sigma/\varepsilon) d\sigma \\
            & \qquad + \sum_{i=2}^\infty \int_t^{t + \vT} A(\sigma_1) + E(\sigma_1/\varepsilon) \int_t^{\sigma_1} \cdots \int_t^{\sigma_{i-1}} A(\sigma_i)+E(\sigma_{i}/\varepsilon) d\sigma_{i} \cdots d\sigma_{1} \notag \\
            & \qquad -I - \int_t^{t + \vT} A(\sigma_1) + \bar{E} d\sigma - \sum_{i=2}^\infty \int_t^{t + \vT} A(\sigma_1) + \bar{E} \int_t^{\sigma_1} \cdots \int_t^{\sigma_{i-1}} A(\sigma_i)+ \bar{E} d\sigma_{i} \cdots d\sigma_{1} \notag \\
\end{split}
\end{equation*}
By hypothesis,
\[
    \int_t^{t+\vT} E(\sigma/\varepsilon) d\sigma = \vT \bar{E}
\]
which implies that 
\begin{equation}
\begin{split} 
    H(t+\vT,t) & =  \sum_{i=2}^\infty \int_t^{t + \vT} A(\sigma_1) + E(\sigma_1/\varepsilon) \int_t^{\sigma_1} \cdots \int_t^{\sigma_{i-1}} A(\sigma_i)+E(\sigma_{i}/\varepsilon) d\sigma_{i} \cdots d\sigma_{1} \notag\\
    & \quad -  \sum_{i=2}^\infty \int_t^{t+\vT} A(\sigma_1) + \bar{E}\int_t^{\sigma_1} \cdots \int_t^{\sigma_{i-1}} A(\sigma_i)+\bar{E} d\sigma_{i} \cdots d\sigma_{1} \notag 
\end{split} 
\end{equation}
Defining 
\begin{equation} \label{eq.define.alpha}
    \alpha \equiv \underset{t\geq 0}{\sup}\left(\max\left( \|A(t) +  \bar{E}\|, \|A(t) +E(t/\varepsilon)\|\right)\right)
\end{equation}
a bound for $H(t+\vT, t)$ is computed
\begin{equation} \label{eq.bound.H}
    \|H(t+\vT, t)\| \leq  2 \big(e^{\vT \alpha} - 1 - \vT \alpha \Big)
\end{equation}
Noting that $\Phi_{E} = \Phi_{\bar{E}} + H$, $\Delta v$ is expressed 
\begin{align}
\begin{split} \label{eq.deltaV}
    \Delta v(z, t+\vT, t) & = z^T(t)\Big( \Phi^T_{\bar{E}}(t+\vT, t)Q(t+\vT)\Phi_{\bar{E}}(t+\vT, t) - Q(t)\Big)z(t) \\
& \quad + z^T(t)\Big( \Phi_{\bar{E}}^T(t +\vT, t)Q(t+\vT) H(t+\vT, t) + H^T(t+\vT,t)Q(t+\vT)\Phi_{\bar{E}}(t+\vT,t) \\
& \quad + H^T(t+\vT,t)Q(t+\vT,t)H(t+\vT,t) \Big)z(t)
\end{split}
\end{align}
The task now is to compute an upper bound for $\Delta v(z, t+\vT,t)$ and show that this bound is negative if $\varepsilon$ is sufficiently small.  Several well-known relationships that are consequences of \eqref{eq.v}, \eqref{eq.vdot}, and uniform exponential stability of \eqref{eq.linE.bar} are utilized (see for example \cite{rugh.book} pages 101 and 117, or \cite{brockett.book}, page 202).  Namely,
\begin{align}
     \|Q(t)\| &\leq \rho     \label{eq.bound.rho} \\
    \|\Phi_{\bar{E}}(t,t_o)\| &\leq \sqrt{\rho /\eta} e^{-\frac{\mu}{2\rho}(t-t_o)}  \label{eq.bound.phi} \\
    v(x(t),t) &\leq e^{-\frac{\mu}{\rho}(t-t_o)} v(x(t_o),t_o) \label{eq.bound.vexp}
\end{align}
for $t \geq t_o$.

To compute an upper bound for the first term on the right-hand side of \eqref{eq.deltaV} we note that if $x(t) = z(t)$ is chosen as the initial condition of \eqref{eq.linE.bar} at time $t$, then 
\begin{equation*}
    z^T(t)\Big( \Phi^T_{\bar{E}}(t+\vT, t)Q(t+\vT)\Phi_{\bar{E}}(t+\vT, t) - Q(t)\Big) z(t)  = v(x(t+\varepsilon T),t+\varepsilon T) - v(x(t),t)
\end{equation*}
From \eqref{eq.bound.vexp} and \eqref{eq.v},
\begin{align*}
\begin{split} 
    v(x(t+\varepsilon T),t+\varepsilon T) - v(x(t),t) &\leq (e^{-\mu\varepsilon T/\rho }-1)v(x(t),t) \\
         &\leq \rho (e^{-\mu \varepsilon T/\rho}-1)\|x(t)\|^2
\end{split}
\end{align*}
Thus,
\begin{equation} \label{eq.bound3}
    z^T(t)\Big( \Phi^T_{\bar{E}}(t+\vT, t)Q(t+\vT)\Phi_{\bar{E}}(t+\vT, t) - Q(t)\Big) z(t) \leq \rho (e^{-\mu \varepsilon T/\rho}-1)\|z(t)\|^2
\end{equation}
Combining \eqref{eq.bound.H}, \eqref{eq.bound.rho}, \eqref{eq.bound.phi}, and \eqref{eq.bound3} yields the desired upper bound 
\begin{equation} \label{eq.v2}
    \Delta v(z,t+\vT, t) \leq \left(\rho (e^{-\mu \varepsilon T/\rho}-1) + 4\rho(\sqrt{\rho /\eta}%
     e^{-\frac{\mu \varepsilon T}{2\rho}})(e^{\vT\alpha} -1 -\vT\alpha) %
     + 4\rho(e^{\vT\alpha} -1 -\vT\alpha)^2\right)\|z(t)\|^2
\end{equation}
Defining the continuously differentiable function $g(\varepsilon,x)$ to be the right-hand side of \eqref{eq.v2}, it can be shown that $g(0,z)=0$ and $\frac{\partial}{\partial \varepsilon}g(0, z)= -\mu T\|z\|^2<0$.  Thus since $g(\varepsilon, z) \rightarrow \infty$ as $\varepsilon \rightarrow \infty$, there exists $\varepsilon^*$ such that $g(\varepsilon^*,z)=0$ and $g(\varepsilon,z) <0$ for all $\varepsilon \in (0, \varepsilon^*)$ and $z\neq 0$. Thus $\Delta v(z, t+\vT,t) <0$ for all $\varepsilon \in (0, \varepsilon^*)$ and $z\neq 0$.  \\

\noindent To show that negative-definiteness of $\Delta v(z, t+\vT,t)$ is sufficient to establish stability of \eqref{eq.linE.t}.  Choose $\varepsilon$ and $\gamma>0$ that satisfy
\[
    \Delta v(z, t_o+\vT,t_o) = v(z(t_o+\vT), t_o+\vT) - v(z(t_o), t_o) \leq -\gamma \|z(t_o)\|^2
\]
for all $t_o$.  From \eqref{eq.v}, $v(z(t_o),t_o) \leq \rho \|z(t_o)\|^2$, which implies that   
\[
    v(z(t_o+\vT), t_o+\vT) -v(z(t_o),t_o) \leq -(\gamma/\rho) v(z(t_o),t_o)
\]
Thus
\[
    v(z(t_o+\vT), t_o+\vT) \leq (1-\gamma/\rho)v(z(t_o),t_o)
\]
Repeating this argument yields
\[
    v(z(t_o+k\vT), t_o+k\vT) \leq (1-\gamma/\rho)^k v(z(t_o),t_o)
\]
for any positive integer $k$. Thus $v(z(t_o +k\vT),t_o +k\vT) \rightarrow 0$ as $k \rightarrow \infty$ which implies that $z(t_o+k\vT) \rightarrow 0$ as $k \rightarrow \infty$.  Since the limiting behavior is valid for any $t_o$, uniform exponential stability of \eqref{eq.linE.t} is established.
\hfill $\blacksquare$



\begin{thebibliography}{10}

\bibitem{skufca-bollt}
J.~D. Skufca, and E. Bollt, ``Communication and Synchronization in Disconnected Networks with Dynamic Topology: Moving Neighborhood Networks," nlin.CD/0307010, \emph{Mathematical Biosciences and Engineering (MBE)}, 1 (2) 347--359 (2004). 

\bibitem{Huygens}
C.~Hugenii, Horoloquium Oscilatorium, Apud F. Muguet, Parisiis, 1673.

\bibitem{classical}
I.~I. Blekman, \emph{Synchronization in Science and Technology}, Nauka, Moscow, 1981 (in Russian); ASME Press, New York, 1988 (in English).

\bibitem{Winfree}
L. Glasss and M.~C. Mackey, {\it From Clocks to Chaos: The Rythms of Life,} Princeton U. Press (1988).

\bibitem{heart1}
J.~Honerkamp, ``The heat as a system of coupled nonlinear oscillators," \emph{J. Math. Bio.}, 19, 69--88 (1983).

\bibitem{heart2}
V.~Torre, ``A theory of synchronization of two heart pacemaker cells," \emph{J. Theor. Bio.}, 61, 55--71 (1976).

\bibitem{heart3}
M.~R.~Guevara, A.~Shrier, and L.~Glass. ``Phase-locked rhythms in periodically stimulated heart cell aggregates,'' \emph{American Journal of Physiology}, 254 (\emph{Heart Circ. Physiol.} 23), H1--H10 (1988).

\bibitem{strogatz1}
R.~Mirollo and S.~Strogatz, ``Synchronization of pulsed-coupled biological oscillators," SIAM J. Appl. Math 50, 6, 1645-1662 (1990).

\bibitem{flies}
 J.~Buck and E.~Buck, ``Synchronous fireflies," \emph{Sci. Am.} 234, 74 (1976).
 
\bibitem{strogatz2}
S.~Strogatz, {\it Sync: The Emerging Science of Spontaneous Order,}  Hyperion, (2003).

\bibitem{gates}
J.~J.~Collins and I.~Stewart, ``Coupled nonlinear oscillators and the symetries of animal gaits," \emph{Science} 3, 3, 349--392 (1993).

\bibitem{schiff}
T.~I. Netoff and S.~J. Schiff, ``Decreased Neuronal Synchronization During Experimental Seizures,'' \emph{Journal of Neuroscience}, 22: 7297--7307, 2002.

\bibitem{postnov}
E. Mosekilde, Y. Maistrenko, and D. Postnov, {\it Chaotic Synchronization: Applications to Living Systems,} World Scientific Nonlinear Science Series A (2002).
 
\bibitem{Roy}
G.~D. Van Wiggeren and R. Roy, \emph{Science} 279 (1998) 1198; R. Roy, K.~S. Thornburg Jr., \emph{Phys. Rev. Lett.} 72 (1994) 2009; J. Ohtsubop, ``Feedback Induced Instability and Chaos in Semiconductor Lasers and Their Applications," \emph{Optical Review,} 6 (1), (1999), 1--15.

\bibitem{chemical}
Y. Kuramoto,  \emph{Chemical Oscillations, Waves and Turbulence},  Springer:Berlin (1984).

\bibitem{meteorology}
G.~S. Duane, P.~J. Webster, and J.~B. Weiss, ``Go-occurrence of Northern and Southern Hemisphere blocks as partially synchronized chaos," \emph{J Atmos Sci}, 56 (24), 4183--4205 (1999).

\bibitem{review1}
S. Boccaletti, J. Kurths, G. Osipov, D.~L. Valladares, and C.~S. Zhou , ``The synchronization of chaotic systems," \emph{Physics Reports} 366 (2002), 1--101.

\bibitem{review2}
 A. Pikovsky, M. Rosemblum, and J. Kurths, {\it Synchronization, A Universal Concept in Nonlinear Sciences,} Cambridge University Press:Cambridge (2001). 
 
\bibitem{review3}
G. Chen and Y. Xinghuo, ``Chaos ControlTheory and Applications Series:  Lecture Notes in Control and Information Sciences," Springer Series 292 (2003).

\bibitem{old1}
H. Fujisaka and T. Yamada, ``Stability theory of synchronized motion in coupled-oscillator systems,'' Prog. Theor. Phys. 69(1) (1983), pp. 32--47. 

\bibitem{old2}
V.~S. Affraimovich, N.~N. Verichev, and and M.~I. Rabinovich, "Stochastic synchronization of oscillation in dissipative systems," Izvestiya Vysshikh Uchebnykh Zavedenii, Radiofizika, 29 (9), pp. 1050–-1060 (1986).

\bibitem{old3}
L.~M. Pecora and T.~L. Carroll, ``Synchronization in Chaotic Systems," \emph{Phys. Rev. Lett.} 64(8), pp. 825--821  (1990). 

\bibitem{reviewn1}
M.~E.~J. Newman,  ``The structure and function of complex networks," \emph{{SIAM} Review} 45, 167--256 (2003). 

\bibitem{reviewn2}
S.~ Dorogovtsev and J.~F.~F. Mendes, "Evolution of networks," Advances in Physics,  51(4), 1079--1187 (2002).

\bibitem{reviewn3}
R. Albert and A.-L. Barabási, ``Statistical mechanics of complex networks," \emph{Reviews of Modern Physics} 74 (1), pp. 47--97 (2002). 

\bibitem{reviewn4}
A.-L. Barabási and E. Bonabeau, ``Scale-Free Networks," \emph{Scientific American} 288, pp. 60--69 (2003).

\bibitem{reviewn5}
D.~J. Watts, {\it Six Degrees: The Science of a Connected Age,} Norton:New York (2003).

\bibitem{net1}
M. Barahona and L.~M. Pecora, "Synchronization in small-world systems," \emph{Phys. Rev. Letters,} 89(5), nlin.CD/0112023 (2002).


\bibitem{net2}
J.~Jost and M.~P. Joy, "Spectral properties and synchronization in coupled map lattices," \emph{Phys. Rev. E} 65, 016201 = nlin.CD/0110037  (2002).


\bibitem{net3}
S.~Jalan and R.~E. Amritkar, "Self-organized and driven phase synchronization in coupled map scale free networks," nlin.AO/0201051.


\bibitem{net4}
H. Hong, M.~Y. Choi, and B.~J.~Kim, "Synchronization on small-world networks," cond-mat/0110359.


\bibitem{net5}
P.~Garcia, A.~Parravano, M.~G. Cosenza, J.~Jimenez, and A.~Marcano, "Coupled map networks as communication schemes," nlin.CD/0201042.


\bibitem{net6}
D.-S.~ Lee, "Synchronization transition in scale-free networks: clusters of synchrony", cond-mat/0410635.

\bibitem{net7}
T. Nishikawa, A.~E. Motter, Y.-C. Lai and F.~C. Hoppensteadt, "Heterogeneity in oscillator networks: Are smaller worlds easier to synchronize?" cond-mat/0306625.

\bibitem{time1}
T.~Stojanovski, L.~Kocarev, U.~Parlitz and R.~Harris, ``Sporadic driving of dynamical systems,"  \emph{Phys. Rev. E,} 55 (4), pp. 4035--4048 (1997).

\bibitem{time2}
 J. Ito amd K. Kaneko, ``Spontaneous structure formation in a network of chaotic units with variable connectionstrengths,"   \emph{Phys. Rev. Lett.,} 88 (2002) 028701.
  
\bibitem{time3}
D.~H. Zanette and A.~S. Mikhailov, "Dynamical systems with time-dependent coupling: clustering and critical behavior", \emph{Physica D}, 194,  pp. 203--218 (2004).

\bibitem{Kataoka}
N. Kataoka and K. Kaneko, ``Dynamical networks in function dynamics," \emph{Physica D}, 181, pp. 235--251  (2003).

\bibitem{syncsymb}
T.~Stojanovski, L. Kocarev, and R. Harris , ``Applications of symbolic dynamics in chaos synchronization,"
\emph{{IEEE} Trans. Circuits and Systems I: Fundamental Theory and Applications}, 44(10), (1997).

\bibitem{ned1}
S.~D. Pethel, N.~J. Corron, Q.~R. Underwood, and K. Myneni, ``Information flow in chaos synchronization:  Fundamental tradeoffs in precision, delay, and anticipation,'' \emph{Phys. Rev. Lett.}, 90, 254101 (2003).

\bibitem{ned2}
N.~J. Corron, S.~D. Pethel, and K.~Myneni, ``Synchronizing the information content of a chaotic map and  flow via symbolic dynamics,'', \emph{Phys. Rev. E}, 66, 036204 (2002).

\bibitem{me1}
E. M. Bollt, ``Review of chaos communication by feedback Control of symbolic dynamics,'' \emph{Int. J. Bifurcation and Chaos} 13(2), pp. 269--285 (2003).

\bibitem{Hayes}
S. Hayes, C. Grebogi, and E. Ott, ``Communicating with chaos," \emph{Phys. Rev. Lett.} 70(20), pp. 3031--3034 (1993).

\bibitem{liberzon.morse.csm1999}
D.~Liberzon and A.~S. Morse, ``Basic problems in stability and design of
  switched systems,'' {\em IEEE Control Systems}, vol.~5, no.~19, pp.~59--70,
  1999.

\bibitem{liberzon.book.2003}
D.~Liberzon, {\em Switching in Systems and Control}.
\newblock Boston: Birkhauser, 2003.

\bibitem{wicks.etal.EJC.1998}
M.~Wicks, P.~Peleties, and R.~DeCarlo, ``Switched controller synthesis for the
  quadratic stabilization of a pair of unstable linear systems,'' {\em European
  Journal of Control}, vol.~4, pp.~140--147, 1998.

\bibitem{bentsman.etal.TAC.1987}
R.~L. Kosut, B.~D.~O. Anderson, and I.~M.~Y. Mareels, ``Stability theory for
  adaptive systems: {M}ethod of averaging and persistency of excitation,'' {\em
  {IEEE} Trans. Automat. Contr.}, vol.~1, no.~32, pp.~26--34, 1987.

\bibitem{bellman.etal.TAC.1985}
R.~Bellman, J.~Bentsman, and S.~M. Meerkov, ``Stability of fast periodic
  systems,'' {\em {IEEE} Trans. Automat. Contr.}, vol.~3, no.~30, pp.~289--291,
  1985.

\bibitem{tokarzewsi.IJSC.1987}
J.~Tokarzewski, ``Stability of periodically switched linear systems and the
  switching frequency,'' {\em International Journal of Systems Science}, vol.~4,
  no.~18, pp.~697--726, 1987.

\bibitem{aeyels.peuteman.automatica.1999}
D.~Aeyels and J.~Peuteman, ``On exponential stability of nonlinear time-varying
  differential equations,'' {\em Automatica}, vol.~35, pp.~1091--1100, 1999.

\bibitem{aeyels.peuteman.TAC.1998}
D.~Aeyels and J.~Peuteman, ``A new aymptotic stability criteria for nonlinear
  time-variant differential equations,'' {\em {IEEE} Trans. Automat. Contr.},
  vol.~43, no.~7, pp.~968--971, 1998.

\bibitem{pecora.carroll.PRL.98}
L.~M. Pecora and T.~L. Carroll, ``Master stability function for synchronized
  coupled systems,'' {\em Physical Review Letters}, vol.~80, no.~10, pp.~2109
  -- 2112, 1998.

\bibitem{Barahona} M. Barahona and L. Pecora, Physical Review Letters, vol 89; 5, 2002.

\bibitem{Fink} K. Fink, G. Johnson, T. Carrol, D. Mar, L. Pecora,  Physical Review E, vol. 61;5, 2000.

\bibitem{pecora.etal.IJBC.2000}
L.~Pecora, T.~Carroll, G.~Johnson, D.~Mar, and K.~S. Fink, ``Synchronization
  stability in coupled oscillator arrays: {S}olution for arbitrary
  configurations,'' {\em International Journal of Bifurcation and Chaos},
  vol.~10, no.~2, pp.~273 –-- 290, 2000.

\bibitem{rugh.book}
W.~J. Rugh, {\em Linear Systems Theory, Second Edition}.
\newblock Upper Saddle River, NJ: Prentic Hall, 1996.

\bibitem{brockett.book}
R.~W. Brockett, {\em Finite Dimensional Linear Systems}.
\newblock New York, NY: John Wiley and Sons,, 1970.

\bibitem{desai.etal.TRA.2001}
J.~P. Desai,  J.~P.~Ostrowski, and V.~Kumar, ``Modeling and Control of Formations of Nonholonomic Mobile Robots",
{\em {IEEE} Trans. Robotics and Automation}, vol.~17, no.~6, pp.~905 -- 908, 2001.

\bibitem{fax.murry.TAC.2004}
J.~A.~Fax and R.~M.~Murry,``Information Flow and Cooperative Control of Vehicle Formations",
{\em {IEEE} Trans. Automatic Control}, vol.~49, no.~9, pp.~1465 -- 147, 2004.

\bibitem{roberson.stilwell.ACC.05}
D.~G.~Roberson and D.~J.~Stilwell,  
``Control of an Autonomous Underwater Vehicle Platoon with a Switched Communication Network",  
{\em Proceedings of the American Control Conference}, Portland, {OR}, 2005.


\end{thebibliography}

\end{document}